\documentclass{elsart}

\usepackage{graphics}
\usepackage{graphicx}

\usepackage{amssymb,amsmath}
\newcommand{\mgb}{MgB$_2$}
\begin{document}

\begin{frontmatter}



\title{Microwave response of a cylindrical cavity made of bulk \mgb\ superconductor}

\author[label1]{A. Agliolo Gallitto},
\author[label1]{G. Bonsignore},
\author[label1,cor1]{M. Li Vigni},
\author[label2]{G. Giunchi},
\author[label3]{Yu. A. Nefyodov}
\address[label1]{CNISM and Dipartimento di Scienze Fisiche e Astronomiche,
Universit\`{a} di Palermo, via Archirafi 36, 90123 Palermo, Italy}
\address[label2]{EDISON S.p.A, Divisione Ricerca e Sviluppo, foro Buonaparte 31,
20121 Milano, Italy}
\address[label3]{Institute of Solid State Physics, Chernogolovka, Moscow District 142432, Russia}
\corauth[cor1]{Tel.:+39 0916234208; fax: +39 0916162461; e-mail:
livigni@fisica.unipa.it}
\begin{abstract}
We report on the microwave properties of a resonant cylindrical
cavity made of bulk \mgb\ superconductor, produced by the reactive
liquid Mg infiltration process. The frequency response of the
cavity has been measured in the range $5\div 13$~GHz. Among the
various modes, the TE$_{011}$, resonating at 9.79~GHz, exhibits
the highest quality factor. For this mode, we have determined the
temperature dependence of the quality factor, obtaining values of
the order of $10^5$ in the temperature range $4.2\div 30$~K. The
values of the surface resistance deduced from the measurements of
the quality factor agree quite well with those independently
measured in a small sample of \mgb\ extracted from the same
specimen from which the cavity has been obtained.
\end{abstract}


\end{frontmatter}

\section{Introduction}
Superconducting materials, because of their low surface
resistance, are recommended for manufacturing many passive
\emph{mw} devices~\cite{lanc,hein}. Among the various devices, the
superconducting resonant cavity is one of the most important
application in the systems requiring high selectivity in the
signal frequency, such as filters for communication systems
\cite{pand}, particle accelerators \cite{padam,collings},
equipment for material characterization at \emph{mw} frequencies
\cite{lanc92,zhai}.

Since the discovery of the high-temperature superconductors (HTS),
several attempts have been done to assemble \emph{mw} cavities
made of bulk HTS~\cite{pand,lanc92,zaho}; however, limitations in
their performance were encountered. Firstly, grain boundaries in
these materials are weakly coupled giving rise to reduction of the
critical current and/or nonlinear effects, which worsen the device
performance~\cite{golo}; furthermore, the process necessary to
obtain bulk HTS in a performing textured form is very elaborated.
For these reasons, in several applications, most of the
superconducting cavities are still manufactured with Nb, requiring
liquid He as refrigerator \cite{padam,zhai,Trunin}.

Since the discovery of superconductivity at 39~K in
\mgb~\cite{naga}, several authors have indicated this material as
promising for technological
applications~\cite{collings,bugo,HeinProc}. The advantages of
using MgB$_2$ for this purpose are several. The simple chemical
composition makes the synthesis process easy, reducing the
production cost in comparison with HTS. The relatively high
critical temperature allows to maintain the compound in the
superconducting state by using modern closed-cycle cryocoolers,
not requiring liquid He as thermal bath. A further interesting
property of \mgb\ bulk samples is the high connectivity of the
superconducting grains. Indeed, contrary to oxide HTS, bulk \mgb\
can be used in the polycrystalline form without a significant
degradation of its critical current
\cite{bugo,HeinProc,larbalestier}. This property has been ascribed
to the large coherence length, which makes the material less
susceptible to structural defects like grain
boundaries~\cite{Rowell}. Due to these amazing properties, MgB$_2$
has been recommended for manufacturing \emph{mw}
cavities~\cite{collings,Tajima}, and investigation is carried out
to test the potential of different MgB$_2$ materials for this
purpose.

In this paper, we report on the \emph{mw} properties of the first
cylindrical cavity made of bulk \mgb~\cite{SUST-cavity}. The
material has been produced by the reactive liquid Mg infiltration
technology~\cite{brevetto,giun03}; it exhibits a critical
temperature $T_c\approx38.5$~K. The frequency response of the
cavity has been measured in the range $5\div 13$~GHz. The quality
factor of the cavity for the TE$_{011}$ mode has been investigated
as a function of the temperature, from $T=4.2$~K up to
$T\approx150$~K. From the $Q$ values, we have determined the
temperature dependence of the \emph{mw} surface resistance of the
\mgb\ material; these results have been compared with those
independently obtained in a small plate-like sample of \mgb\
extracted from the same specimen used for the cavity.

\section{Cylindrical cavities: theoretical aspects}
As it is well known, an important valuation index to determine the
performance of a resonant cavity is the quality factor $Q$,
defined as
\begin{equation}
    Q=2\pi\frac{energy\ stored\ in\ the\ cavity}{energy\ lost\ per\
    cycle}\,.
\end{equation}
In an empty resonant cavity, essentially two causes contribute to
the energy losses: conductor losses, associated to the conduction
currents in the cavity walls; radiation losses, due to power that
leaves the cavity through the holes for the coupling between the
resonator and the external circuit. Therefore, the overall or
loaded $Q$, denoted by $Q^L$, can be defined by
\begin{equation}
\label{equ:Q-L} \frac{1}{Q^L}=\frac{1}{Q^U} + \frac{1}{Q^R}\,,
\end{equation}
where $Q^R$ is due to the radiative losses and $Q^U$, the
so-called unloaded $Q$, includes only the conductor
losses.\\
$Q^U$ is given by
\begin{equation}
\label{equ:Q-U} Q^U=\frac{\Gamma }{R_s}\,,
\end{equation}
where $R_s$ is the surface resistance of the material; $\Gamma$ is
the geometry factor, it depends on the shape and dimensions of the
cavity and the specific resonant mode.

For a cavity coupled to the external circuit through two lines,
$Q^U$ can be deduced, from the measured $Q^L$, by taking into
account the coupling coefficients, $\beta_1$ and $\beta_2$, for
both the coupling lines; these coefficients can be calculated by
directly measuring the reflected power at each line, as described
in Chap. IV of Ref.~\cite{lanc}. Thus, $Q^U$ can be calculated as
\begin{equation}
Q^U = (1+\beta_1+\beta_2)Q^L\,.
\end{equation}
It is worth noting that $Q^R$ can be made large by weakly coupling
the excitation and detection lines. In this case, $\beta_1 +
\beta_2 \ll 1$ and $Q^U \approx Q^L$.

As it is well known, resonant cylindrical cavities can support
both TE$_{lmn}$ and TM$_{lmn}$ modes; the subscripts \emph{l},
\emph{m} and \emph{n} refer to the number of half-wavelength
variations in the standing-wave pattern in $\phi$, $r$ and $z$
directions, respectively. The resonant frequencies of a
cylindrical cavity resonating in the TE$_{lmn}$ or TM$_{lmn}$
modes are given by~\cite{lanc}
\begin{equation}
\label{equ:f-TE} f^{TE}_{lmn} = \frac{1}{2 \pi\sqrt{\epsilon\mu}}
\sqrt{\left(\frac{n\pi}{d}\right)^2+\left(\frac{z_{lm}^{\prime}}{a}\right)^2}\,,
\end{equation}
\begin{equation}\label{equ:f-TM}
f^{TM}_{lmn} = \frac{1}{2 \pi\sqrt{\epsilon\mu}}
\sqrt{\left(\frac{n\pi}{d}\right)^2+\left(\frac{z_{lm}}{a}\right)^2}\,,
\end{equation}
where $\mu$ and $\epsilon$ are the permeability and dielectric
constant of the medium filling the cavity; $a$ and $d$ the radius
and length of the cavity; $z_{lm}$ and $z_{lm}^{\prime}$  are  the
\emph{m}th zeros of the Bessel function of order \emph{l} and of
its first derivative, respectively.

Among the various resonant modes, the ones more extensively used
are the TE$_{01n}$, which give a reasonably high quality factor.
In the TE$_{01n}$ modes, the wall currents are purely
circumferential and no currents flow between the end plates and
the cylinder. This implies that no electrical contact between
plates and cylinder is required, making the cavity assembly easy.
The TE$_{01n}$ modes are degenerate in frequency with the
TM$_{11n}$ modes and this should be avoided to have a well defined
field configuration; however, this degeneracy can be removed
introducing a small gap between cylinder and lids. This shifts the
resonant frequency of the TM$_{11n}$ modes downwards, leaving the
TE$_{01n}$ modes nearly unperturbed. The unloaded quality factor
of a cylindrical cavity resonating in the TE$_{01n}$ mode is given
by
\begin{equation}\label{equ:Q-TE01n}
Q_{01n}^U = \frac{\Gamma_{01n}}{R_s}\,,
\end{equation}
with
\begin{equation*}\label{equ:GAMMA-TE01n} \Gamma_{01n} =
\sqrt{\frac{\mu}{\epsilon}} \frac{\left[(z_{01}^{\prime}d)^2 + (n
\pi a)^2\right]^{3/2}}{2(z_{01}^{\prime})^2 d^3 + 4 n^2\pi^2
a^3}\,,\;\;\; z_{01}^{\prime} = 3.83170\,.
\end{equation*}

\section{The superconducting \mgb\ cavity}\label{cavity}
The material from which the cavity is made has been produced by
the reactive liquid \mgb\ infiltration technology
(RLI)~\cite{brevetto,giun03}. The RLI process consists in the
reaction of pure liquid Mg and boron powder, sealed in a stainless
steel container with a weight ratio Mg/B over the stoichiometric
value ($\approx 0.55$). The reaction between liquid Mg and B
powder produces a more stable material, with respect to the hot
pressing technique, at temperatures of a few hundred K above the
Mg melting point and at moderate pressure. This procedure allows
one to obtain specimens of high density and dimensions of the
order of tens on centimeters, showing very high mechanical
strength~\cite{giun03,giun-cryo06}.

As the first attempt to apply the MgB$_2$ produced by RLI to the
cavity-filter technology, we have manufactured a simple
cylindrical cavity and have investigated its microwave response in
a wide range of frequencies. All the parts of the cavity, cylinder
and lids, are made of bulk MgB$_2$ material with $T_c \approx
38.5$~K and density $\approx 2.33~\mathrm{g/cm^3}$. The material
has been obtained using commercial crystalline B powder (Starck H.
C., Germany, 99.5\% purity) and pure Mg; in order to get the B
powder to be used in the synthesis process, the original chunks
were mechanically crushed and filtered through a 100 micron sieve.
In particular, the present cylindrical cavity (inner diameter 40
mm, outer diameter 48 mm, height 42.5 mm) was cut by
electroerosion from a thicker bulk MgB$_2$ cylinder, previously
prepared as described in Sec.~4.3 of Ref.~\cite{giun-cryo06},
internally polished to a surface roughness of about 300~nm.

\begin{figure}\label{Fig-cavity}
\centering
\includegraphics[width=0.45\textwidth]{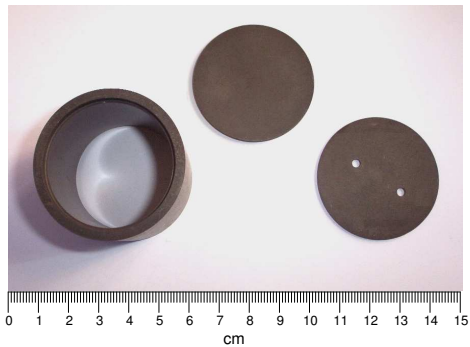}
\caption{Photograph of the cylinder and the lids composing the
bulk MgB$_2$ cavity. The holes in one of the lids are used to
insert the coupling loops. After Ref.\cite{SUST-cavity}.}
\end{figure}

A photograph of the parts, cylinder and lids, composing the
superconducting cavity is shown in Fig.~\ref{Fig-cavity}. The
holes in one of the lids have been used to insert two small loop
antennas which couple the cavity with the excitation and detection
lines. The loop antenna was constructed on the end of each line,
soldering the central conductor to the outer shielding of the
semirigid cable. In order to remove the degeneracy between
TE$_{01n}$ and TM$_{11n}$ modes, we have incorporated a ``mode
trap" in the form of circular grooves (1 mm thick, 2 mm wide)
inside the cylinder at the outer edges.

\section{Experimental results}\label{results}
\begin{figure}[ht]
\centering
\includegraphics[width=7.3cm]{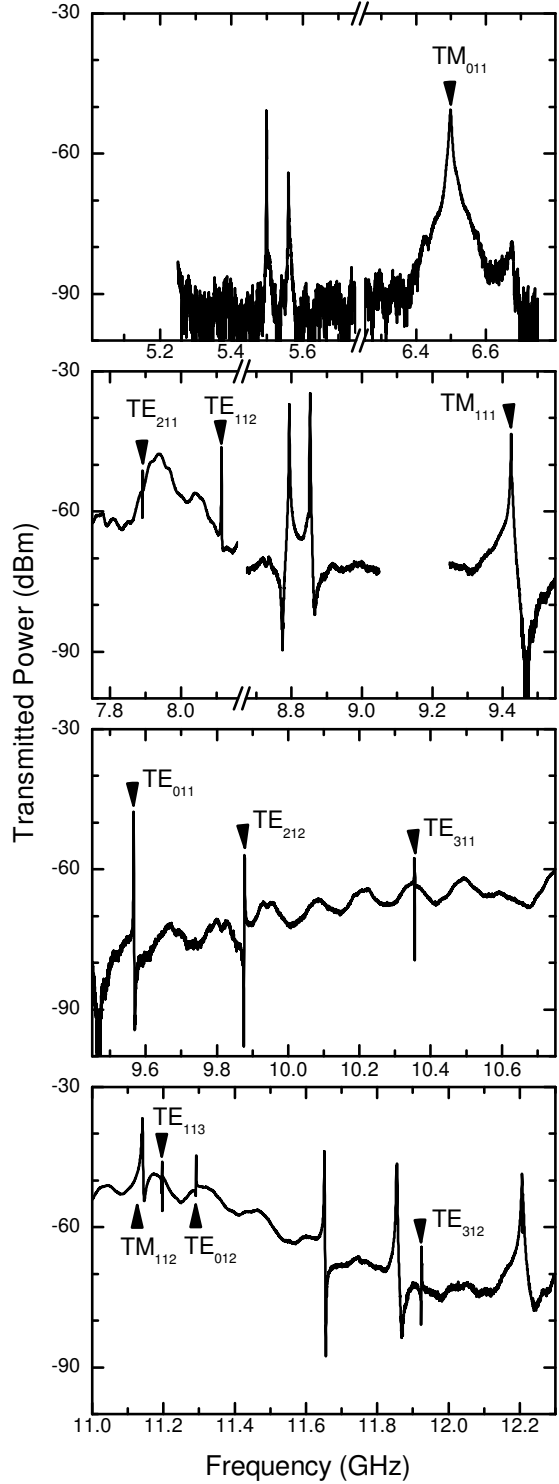}
\caption{Microwave response of the \mgb\ cavity filled by liquid
He in the frequency range 5.3~$\div$~12.3~GHz. In the omitted
ranges, no resonant curves were present.} \label{fig:spettro}
\end{figure}
The resonant cavity has been characterized by measuring its
frequency response in the range $5\div 13$~GHz by an
\textit{hp}-8719D Network Analyzer, using a two-port
configuration. The spectrum of the cavity, measured at $T=4.2$~K,
is shown in Fig.~\ref{fig:spettro}, in four frequency intervals.
The arrows indicate the resonances we have certainly recognized by
Eqs.~(\ref{equ:f-TE}) and (\ref{equ:f-TM}); some peaks show a not
simple structure, probably due to the superposition of different
modes, which hinders to recognize the corresponding mode.

\begin{figure}[t]
  \centering
  \includegraphics[width=7.5cm]{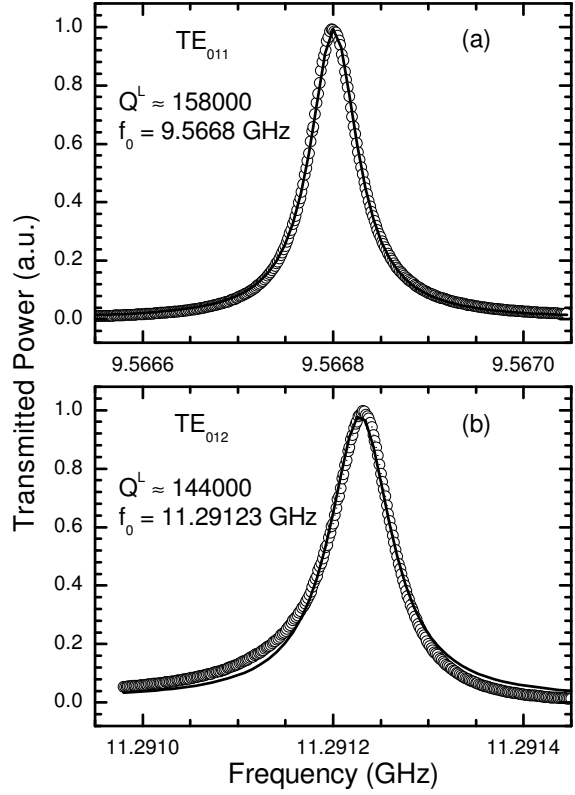}
\caption{Resonant curves, at $T=4.2$~K, corresponding to the
TE$_{011}$ (a) and TE$_{012}$ (b) modes. The lines are Lorentzian
fits of the experimental curves from which we have extracted the
values of $Q^L$ and the central frequency $f_0$.}
\label{fig:TE-curve}
\end{figure}

Among the various modes detected, two of them show the highest
quality factors; at $T=4.2$ K, with the cavity filled by liquid
He, the resonant frequencies of these modes are 9.567 GHz and
11.291 GHz. At room temperature, with the cavity filled by gaseous
He, these resonant frequencies move to 9.79~GHz and 11.54~GHz;
according to Eq.~(\ref{equ:f-TE}), they correspond to the
TE$_{011}$ and TE$_{012}$ modes. Fig.~\ref{fig:TE-curve} shows the
resonant curves, at $T=4.2$~K, corresponding to the TE$_{011}$ (a)
and TE$_{012}$ (b) modes. The lines are Lorentzian fits of the
experimental data, which give for the loaded quality factors
$Q_{011}^L \approx 158000$ and $Q_{012}^L \approx 144000$.

\begin{figure}[t]
  \centering
  \includegraphics[width=7.5cm]{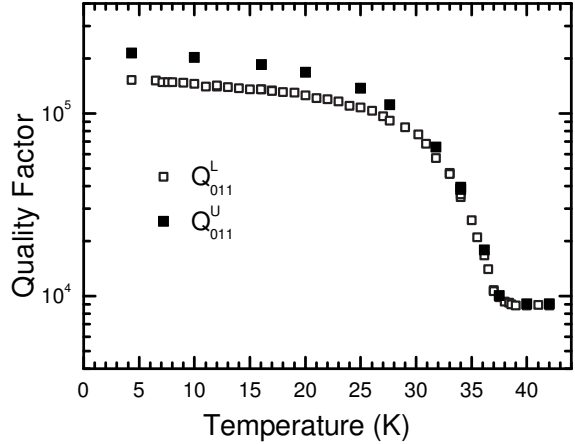}
\caption{Temperature dependence of the loaded- and unloaded-$Q$
values, $Q_{011}^L$ and $Q_{011}^U$, for the TE$_{011}$ mode.}
\label{fig:QvsT-cavity}
\end{figure}

\begin{figure}[htbp]
  \centering
  \includegraphics[width=7.5cm]{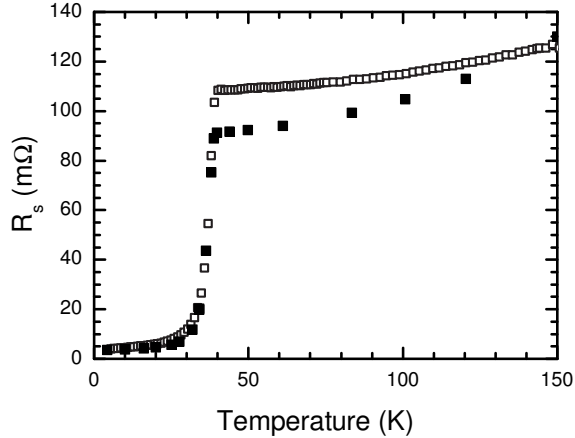}
\caption{Comparison between the temperature dependence of $R_s(T)$
deduced from $Q_{011}^U(T)$ (full symbols) and that obtained at
9.4~GHz in a small \mgb\ sample extracted from the same material
of the cavity (open symbols).} \label{fig:RvsT-cavity}
\end{figure}

The quality factor of the TE$_{011}$ mode has been measured as a
function of the temperature, in the range $4.2\div 150$~K. In
order to calculate the unloaded $Q$-values, we have determined the
coupling coefficients by measuring the power reflected by each
line. The coupling coefficients, $\beta_1$ and $\beta_2$, have
been measured at different values of the temperature; they result
$\approx 0.2$ at $T=4.2$~K and reduce to $\approx 0.05$ when the
material goes to the normal state, at $T \approx 38.5$~K.
Fig.~\ref{fig:QvsT-cavity} shows the temperature dependence of the
loaded and unloaded $Q$ for the TE$_{011}$ mode, $Q_{011}^L$ and
$Q_{011}^U$. At $T=4.2$~K, we have obtained
$Q_{011}^U=2.2\times10^5$; on increasing the temperature, the
quality factor maintains values of the order of $10^5$ up to $T
\approx 30$~K and reduces by a factor of $\approx 20$ at $T=T_c$.
From the values of $Q_{011}^U$, it is possible, using
Eq.~(\ref{equ:Q-TE01n}), to deduce the surface resistance of the
\mgb\ material from which the cavity is made. In order to identify
possible spurious effects in the measurement of the \emph{mw}
response of the cavity, we have measured the \emph{mw} surface
resistance of a small plate-like sample of \mgb\ extracted from
the same specimen from which the cavity has been obtained. These
measurements have been performed by the technique of hot-finger
cavity perturbation~\cite{Trunin}, using a Nb cavity resonating at
$\approx 9.4$~GHz. Fig.~\ref{fig:RvsT-cavity} shows a comparison
between the $R_s(T)$ values deduced from $Q_{011}^U(T)$ (full
symbols) and those obtained in the sample (open symbols). As one
can see, when the cavity is in the superconducting state, the
$R_s(T)$ curves obtained by the two different techniques are
consistent, while for $T>T_c$ they differ by $\approx 10 \%$. This
disagreement may be ascribed to the difficulty of measuring with
high sensitivity the quality factor of the \mgb\ cavity at
$T>T_c$, due to its small value. On the contrary, the sensitivity
of the measurements performed by the technique of hot-finger
cavity perturbation is very high independently of the temperature;
indeed, the Nb cavity is always maintained in the superconducting
state and only the temperature of the sample is changed.

As one can see from Fig.~\ref{fig:TE-curve}b, the resonant curve
for the TE$_{012}$ mode is not very well described by a Lorentzian
function; this problem becomes more significant on increasing the
temperature, making inaccurate the determination of the quality
factor at temperatures higher that $\approx 30$~K. For these
reason, we have not reported here the temperature dependence of
$Q_{012}$. However, we have verified that, also for the TE$_{102}$
mode, the quality factor takes on values of the order of  $10^5$
up to $T\sim 30$~K and reduced by a factor of $\approx 20$ when
the material goes to the normal state.

\section{Discussion and Conclusion}
It is well known that one of the most important application of
superconducting materials is the implementation of \emph{mw}
resonant cavities, which can be used in a large variety of
devices~\cite{lanc,pand,padam,collings,lanc92,zhai}. Although
several studies have been devoted to build \emph{mw} cavities
using HTS, actually most of the superconducting cavities are still
manufactured with Nb, requiring liquid He as refrigerator. The
main problem encountered in using bulk HTS, for this purpose, is
due to the weak coupling of grain boundaries in these materials,
which act as weak links. On the contrary, it has widely been shown
that in bulk \mgb\ samples only a small amount of grain boundaries
act as weak links~\cite{Rowell,Samanta,Khare,agliJ-SC}, hindering
the strong suppression of the critical current with magnetic field
and the generation of nonlinear effects at high input power. For
these reasons, as well as for the relatively high $T_c$, several
authors have recommended \mgb\ as a convenient material to build
\emph{mw} cavity.

Recently, we have investigated the \emph{mw} response of small
plate-like samples of MgB$_2$ prepared by RLI process, in the
linear and nonlinear regimes~\cite{agliJ-SC,agli-EUCAS}. These
studies have highlighted a weak nonlinear response, as well as
relatively small values of the residual surface resistance. In
particular, we have found that the \emph{mw} properties of such
samples improve on decreasing the size of the B powder used in the
synthesis process. Furthermore, bulk samples produced by RLI
maintain the surface staining unchanged for years, without
controlled-atmosphere protection. This useful property is most
likely related to the high density, and consequently high grain
connectivity, achieved with the RLI process, as well as to the
small and controlled amount of impurity phases
\cite{giun-physicaC}. Prompted by these interesting results, we
have built the \emph{mw} resonant cavity using MgB$_2$ produced by
the RLI process~\cite{SUST-cavity}.

The results of Fig.~\ref{fig:QvsT-cavity} show that $Q$ takes on
values of the order of $10^5$ from $T=4.2$~K up to $T\approx
30$~K, a temperature easily reachable by modern closed-cycle
cryocoolers. To our knowledge, these $Q$ values are higher than
those reported in the literature for \emph{mw} cylindrical
cavities manufactured with HTS, both bulk and films
\cite{pand,lanc92,zaho}. So, our results show that MgB$_2$
produced by RLI is a very promising material for building
\emph{mw} resonant cavities. We would remark that this is the
first attempt to realize a superconducting cavity made of bulk
\mgb. In particular, this investigation has been carried out with
the aim to explore the potential of bulk MgB$_2$ materials
prepared by RLI to the \emph{mw}-cavity technology.

The MgB$_2$ material from which the present cavity is made has
been obtained using crystalline B powder with grain mean size
$\approx 100~\mu$m. Our previous studies on bulk MgB$_2$ samples,
obtained by the RLI method, have shown that samples prepared using
microcrystalline B powder ($\approx 1~\mu$m in size) exhibit
smaller residual surface resistance ($\approx 0.5~\mathrm{m}
\Omega$)~\cite{agli-EUCAS}. From Fig.~\ref{fig:RvsT-cavity}, one
can see that the residual surface resistance of the cavity is
$R_s(4.2~\mathrm{K})\approx 3.5~\mathrm{m}\Omega$. So, we infer
that one could improve the quality factor by one order of
magnitude manufacturing the cavity with material produced by
liquid Mg infiltration in micrometric B powder.

Because of the shorter length of percolation of the liquid Mg into
very fine B powder, the production of massive MgB$_2$ samples by
RLI using B powder with size $\approx 1~\mu$m turns out to be more
elaborate. However, work is in progress to improve the preparation
process in order to manufacture large specimens using
microcrystalline B powder. Considering that the \emph{mw} fields
penetrate in a surface layer of the material of the order of the
penetration depth, a more performing \emph{mw} resonant cavity can
be obtained designing a composite structure; in particular, one
can use microcrystalline B powder for the inner part of the cavity
and larger grain-size B powder for the outer part.

In conclusion, we have shown that the RLI process provides a
useful method for assembling high-performance \emph{mw} cavities,
which may have large scale application. We have obtained quality
factors of the order of $10^5$, larger than those reported in the
literature for cavities made of HTS, both bulk and films, in the
same temperature range. Although higher quality factors have been
reported for superconducting cavities made of pure Nb and Nb
alloys~\cite{padam}, Nb cavities must be kept in liquid-He bath;
on the contrary, cavity made of MgB$_2$ can be maintained in the
superconducting state by using close-cycle cryocoolers that can
easily work at temperature of the order of 10~K.


\section*{Acknowledgements}
The authors are very glad to thank M. Bonura for his continuous
interest and helpful suggestions, G. Lapis and G. Napoli for
technical assistance. Work partially supported by the University
of Palermo (grant Coll. Int. Li Vigni, Co.RI 2005).



\begin{thebibliography}{99}


\bibitem{lanc}M. J. Lancaster, \emph{Passive Microwave Device Applications of
High-Temperature Superconductors}, Cambridge University Press
(Cambridge 1997).

\bibitem{hein}M. Hein, \emph{High-Temperature Superconductor Thin Films
at Microwave Frequencies}, Springer Tracts of Modern Physics, vol.
\textbf{155}, Springer (Heidelberg 1999).

\bibitem{pand}H. Pandit, D. Shi, N. H. Babu, X. Chaud, D. A. Cardwell, P. He,
D. Isfort, R. Tournier, D. Mast, and A. M. Ferendeci, Physica
\textbf{C 425} (2005) 44.

\bibitem{padam}H. Padamsee, Supercond. Sci. Technol. \textbf{14} (2001) R28.

\bibitem{collings}E. W. Collings, M. D. Sumption, and T. Tajima,
Supercond. Sci. Technol. \textbf{17} (2004) S595.

\bibitem{lanc92}M. J. Lancaster, T. S. M. Maclean, Z. Wu, A. Porch, P. Woodall,
N. NcN. Alford, IEE Proceedings-H, vol. \textbf{139} (1992) 149.

\bibitem{zhai}Z. Zhai, C. Kusko, N. Hakim, and S. Sridhar, Rev.
Sci. Instrum. \textbf{71} (2000) 3151.

\bibitem{zaho}C. Zahopoulos, W. L. Kennedy, S. Sridhar, Appl. Phys. Lett.
\textbf{52} (1988) 2168.

\bibitem{golo}M. Golosovsky, Particle Accelerators \textbf{351} (1998) 87,
and references therein.

\bibitem{Trunin}M. R. Trunin, Physics-Uspeki \textbf{48} (2005) 979.

\bibitem{naga}J. Nagamatsu, N. Nakagawa, T. Muranaka, Y. Zenitani, and
J. Akimitsu, Nature (London) \textbf{410} (2001) 63.

\bibitem{bugo}Y. Bugoslavsky, G. K. Perkins, X. Qi, L. F. Cohen, and A. D. Caplin,
Nature (London) \textbf{410} (2001) 563.

\bibitem{HeinProc}M. A. Hein, \emph{Proceedings of URSI-GA }(Maastricht 2002);
e-print arXiv:cond-mat/0207226.

\bibitem{larbalestier}D. C. Larbalestier, L. D. Cooley,  M. O. Rikel, A.A. Polyanskii,
 J. Jiang, S. Patnaik, X. Y. Cai, D. M. Feldmann, A. Gurevich, A. A.  Squitieri, M. T. Naus,
 C. B. Eom, E. E. Hellstrom, R. J. Cava, K. A. Regan, N. Rogado, M. A. Hayward,
 T. He, J. S. Slusky, P. Khalifah, K. Inumaru, and M. Haas,
 Nature (London) \textbf{410} (2001) 186.

\bibitem{Rowell} J. M. Rowell, Supercond. Sci.
Technol. \textbf{16} (2003) R17.

\bibitem{Tajima}T. Tajima, \emph{Proceedings of EPAC Conf.} (Paris 2002) 2289.

\bibitem{SUST-cavity}G. Giunchi, A. Agliolo Gallitto, G. Bonsignore, M. Bonura
and M. Li Vigni, Supercond. Sci. Technol. \textbf{20} (2007) L16.

\bibitem{brevetto}EDISON, \emph{patent n. MI2001A000978}.

\bibitem{giun03}G. Giunchi, Int. J. Mod. Phys. \textbf{B 17} (2003) 453.

\bibitem{giun-cryo06}G. Giunchi, G. Ripamonti, T. Cavallin, E.
Bassani, Cryogenics \textbf{46} (2006) 237.

\bibitem{Samanta}S. B. Samanta, H. Narayan, A. Gupta, A. V. Narlikar, T. Muranaka,
and J. Akimtsu, Phys. Rev. \textbf{B 65} (2002) 092510.

\bibitem{Khare}Neeraj Khare, D. P. Singh, A. K. Gupta, Shashawati Sen, D. K.
Aswal, S. K. Gupta, and L. C. Gupta, J. Appl. Phys. \textbf{97}
(2005) 07613.

\bibitem{agliJ-SC}A. Agliolo Gallitto, G. Bonsignore, G. Giunchi, and M. Li Vigni, J.
Supercond. \textbf{20} (2007) 13.

\bibitem{agli-EUCAS}A. Agliolo Gallitto, G. Bonsignore, G. Giunchi, M. Li Vigni, and
Yu. A. Nefyodov, J. Phys.: Conf. Ser. \textbf{43} (2006) 480.

\bibitem{giun-physicaC}G. Giunchi, C. Orecchia, L. Malpezzi, and
N. Masciocchi, Physica \textbf{C 433} (2006) 182.


\end{thebibliography}
\end{document}